# Single-shot quantitative polarization imaging of complex birefringent structure dynamics

Baoliang Ge,[1,2,#] Qing Zhang,[1,#] Rui Zhang,[3] Jing-Tang Lin,[4] Po-Hang Tseng,[4] Che-Wei Chang,[4] Chen-Yuan Dong,[4] Renjie Zhou,[5] Zahid Yaqoob,[2] Irmgard Bischofberger[1] and Peter T. C. So[1,2,6,*]

[1]Department of Mechanical Engineering, Massachusetts Institute of Technology, Cambridge, MA 02139, USA

[2]Laser Biomedical Research Center, Massachusetts Institute of Technology, Cambridge, MA 02139, USA

[3]Department of Physics, The Hong Kong University of Science and Technology, Hong Kong, China

[4]Department of Physics, National Taiwan University, Taipei 106, Taiwan, Republic of China

[5]Department of Biomedical Engineering, The Chinese University of Hong Kong, Shatin, New Territories, Hong Kong, China

[6]Department of Biological Engineering, Massachusetts Institute of Technology, Cambridge, MA 02139, USA

[#]These authors contributed equally to this work

*Corresponding author: ptso@mit.edu

## ABSTRACT

Polarization light microscopes are powerful tools for probing molecular order and orientation in birefringent materials. While a multitude of polarization light microscopy techniques are often used to access steady-state properties of birefringent samples, quantitative measurements of the molecular orientation dynamics on the millisecond time scale have remained a challenge. We propose polarized shearing interference microscopy (PSIM), a single-shot quantitative polarization imaging method, for extracting the retardance and orientation angle of the laser beam transmitting through optically anisotropic specimens with complex structures. The measurement accuracy and imaging performances of PSIM are validated by imaging a rotating wave plate and a bovine tendon specimen. We demonstrate that PSIM can quantify the dynamics of a flowing lyotropic chromonic liquid crystal in a microfluidic channel at an imaging speed of 506 frames per second (only limited by the camera frame rate), with a field-of-view of up to 350×350 μm$^2$ and a diffraction-limit spatial resolution of ∼2 μm. We envision that PSIM will find a broad range of applications in quantitative material characterization under dynamical conditions.

# INTRODUCTION

Many materials, including liquid crystals (LCs) [1], collagen fibers [2], and cytoskeletons [3], are optically anisotropic. The structural anisotropy usually leads to refractive index anisotropy, termed birefringence. Polarization light microscopes are powerful imaging tools to study these birefringent materials [4], since they can visualize the birefringent structures via detecting the light components in different polarization directions. One of the most commonly used polarization light microscopy techniques for characterizing molecular orientational conformation in birefringent specimens is fluorescence polarization-resolved microscopy [5,6], which is based on doping the material with anisometric fluorescent dye molecules that report on the surrounding molecular director field [5,7]. This fluorescence approach can provide high-contrast and high-resolution imaging of birefringent structures, though suffers from several shortcomings. The addition of fluorescent agents may affect the behavior of the underlying molecular matrix unless the doping concentration is very low. The imaging speed and the observation time is limited by the concentration and photobleaching of the fluorescent agents. Finally, implementing fluorescence approaches with confocal detection [7] or light-sheet microscopy [8,9] enables mapping of the molecular director field in three dimensions, however, raster scanning or depth scanning approaches further limit the imaging speed to a few frames per second [7,10,11]. Although recent approaches have achieved millisecond temporal resolution [12,13], the advances in imaging speed has the cost of complicated instrumentation, decreasing its applicability in studies of birefringent structure dynamics.

As a label-free alternative, polarization optical microscopes (POMs) [14] based on crossing a polarizer and an analyzer before and after the specimen, are often used for the imaging of the underlying molecular assembly in birefringent materials. POMs have found applications in studying flow-rate dependent liquid crystal conformational profiles [15,16], analyzing collagen fibers' architectures [2], or measuring the photoelasticity of glass [17]. While qualitative polarization imaging with POMs have been broadly used, the quantitative polarization imaging is also significant for the investigations of optically anisotropic materials. Retardance and orientation angle are the two essential polarization parameters typically used for quantifying the properties and structures of birefringent specimens [14]. However, to resolve these two polarization parameters quantitatively, multiple measurements based on mechanical rotations of the polarizer and analyzer are needed [14] in conventional POMs, which is often too slow to study the dynamics of birefringent materials. To overcome those limitations, LC-Polscope [18,19] has been developed that improved the accuracy, sensitivity, and speed of quantitative polarization imaging by replacing the compensators in conventional POMs with faster electro-optical LC universal retarders [4,20,21]. These improvements make LC-Polscope a powerful tool for studying biological processes such as microtubule reorganization during cell mitosis [22,23]. LC-Polscope has also been used to study dynamical events of birefringent samples such as LC flows [24], but the imaging speed remains restricted by the requirement of multiple measurements to reconstruct the retardance and orientation angle maps.

Driven by the desire to explore fast dynamics in birefringent structures, the development of the next generation of polarization light microscopes that can rapidly quantify the retardance and the orientation angle is essential. Recent approaches combine polarization light microscopy with quantitative phase imaging (QPI) [25], where the scattered optical complex field is retrieved with the implementation of off-axis interferometry. With an off-axis QPI system, both the amplitude and the phase of the transmitted light can be obtained in a single snapshot [26]. Incorporating off-axis QPI into polarization light microscopy allows one to reduce the number of measurements for quantitative retrieval of polarization parameters [6,27–29], but these approaches still require multiple measurements. We had first overcome this limitation in a single-shot quantitative polarization imaging technique utilizing shearing interferometry and a novel retrieval algorithm [30]. However, the Wollaston prism used in this system severely limits the image field-of-view

(FOV) to a narrow rectangle. Moreover, the use of Wollaston prisms requires low numerical aperture objectives, which reduces the spatial resolution. Finally, the retardance and orientation angle retrieval algorithm amplifies the measurement noise and deteriorates the imaging sensitivity. Hindered by these limitations, this single-shot quantitative polarization imaging method is still not suitable for studying dynamics of birefringent samples with complex structures.

Here we propose polarized shearing interference microscopy (PSIM) that overcomes all these limitations. We utilize a diffraction grating with polarizer sheets in the Fourier plane instead of a Wollaston prism, which enables single-shot quantitative polarization imaging with large FOV imaging (up to 350×350 μm$^2$) with high spatial resolution (~2 μm). A new polarization parameter retrieval algorithm is developed that avoids noise amplification and improves the sensitivity. The introduction of a supercontinuum laser source reduces image speckle noise, and the use of a fast CMOS camera improves the acquisition speed. We show the capabilities of PSIM by imaging the flow of a lyotropic chromonic liquid crystal (LCLC) in a microfluidic channel at an imaging speed of 506 fps. LCLCs are biocompatible water-based materials composed of self-assembled cylindrical aggregates [31–34]. Despite being a good candidate for controlling assembly of particles and biomaterials in microfluidic channels [35–37], the flow properties of LCLCs are currently poorly understood [37,38]. We show that the evolution of the retardance and orientational angle maps of the flowing LCLC can be quantified with PSIM.

## TECHNIQUE

**Polarized shearing interference microscopy (PSIM).** A supercontinuum laser (Fianium SC-400) generates a broadband, spatially uniform illumination beam that is coupled to a single-mode optical fiber, as shown in Fig. 1. The beam is collimated and transmits through a bandpass filter centered at 633 nm with 10 nm bandwidth. A linear polarizer (LP1) and a quarter wave plate (QWP1) are crossed at an angle of 45° to generate a circularly polarized illumination. The scattered light is collected by an objective (Olympus UPLFLN10X2, NA = 0.3, 10X) and collimated by a tube lens (TL). The beam, which bears the polarization information of the sample, transmits through another quarter wave plate (QWP2) and then is separated into multiple orders by a 100 line-pair per millimeter (LP) diffraction grating. The grating is positioned at the conjugated plane of the sample plane. A 4f system is positioned after the diffraction grating to relay the beams. On the Fourier plane, a mask is placed that lets only the +1$^{st}$ order and –1$^{st}$ order beams pass. Two polarizer sheets are placed on the mask: For the +1$^{st}$ order, the direction of the polarization sheet is 45° to the slow axis of QWP2, for the –1$^{st}$ order, the direction of the polarization sheet is –45° to the slow axis of QWP2. Another linear polarizer (LP2) with polarization direction 45° to both polarizer sheets is placed in front of the camera to produce interference between the two orders. The interferogram is recorded by a CMOS camera (Optronics CP80-4-M-500, full frame 2304 × 1720 pixels, pixel size 7 × 7 μm$^2$), whose maximum frame rate is 506 frames per second (fps). The imaging speed of PSIM is limited only by the camera frame rate.

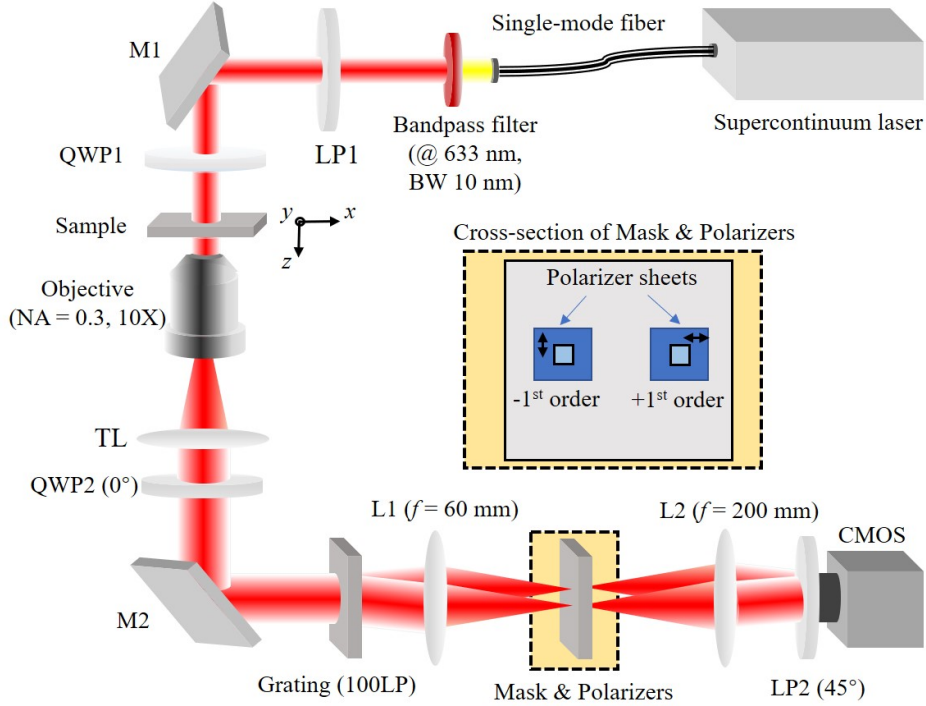

Figure 1. System design of polarized shearing interference microscopy. LP1, LP2, linear polarizers; M1, M2, mirrors; QWP1, QWP2, quarter wave plates; TL, tube lens; L1, L2, lenses. The *z*-axis is the direction of the optical axis, and the *x-y* plane is the sample plane. The zoomed region denotes the location of the masks and polarizer sheets on the Fourier plane.

**Polarization retrieval algorithm.** We have developed a new algorithm based on digital holography to quantitatively retrieve the 2D polarization parameters of birefringent samples. As a demonstration of this algorithm, we recover the polarization parameters of a crystallized Orange II fiber from its interferogram, as shown in Fig. 2(a). In the interferogram, fringes appear only in regions with high birefringence. We perform a two-dimensional Fourier transform to the interferogram and show the logarithm of the 2D spectrum in decibel (dB), as shown in Fig. 2(b), which reveals three orders on the Fourier domain ($-1^{st}$, $0^{th}$, $+1^{st}$ order). By extracting the $+1^{st}$ order with a circular linear filter and shifting it to the center of the Fourier plane, we can map the amplitude $E$ and the phase $\phi$ of the light field after an inverse Fourier transform. The $0^{th}$ order gives access to the amplitude of the direct current (DC) term $A$. The retardance $\Delta$ can be calculated as

$$\Delta = \sin^{-1}\left(\frac{2E}{A}\right), \tag{1}$$

and the distribution of the orientation angle $\varphi$ is calculated as

$$\varphi = \frac{1}{2}\phi, \tag{2}$$

The retardance $\Delta$ and the orientation angle distribution $\varphi$ are decoupled into the measured amplitude $E$ and the measured phase $\phi$, respectively, as shown in Figs. 2(c) and (d). The derivation of the polarization parameter retrieval, and the preparation of the Orange II samples, are discussed in the Methods section. Besides, the calibration approaches used to eliminate the background noise and to improve the imaging accuracy are elaborated in the Supporting Information.

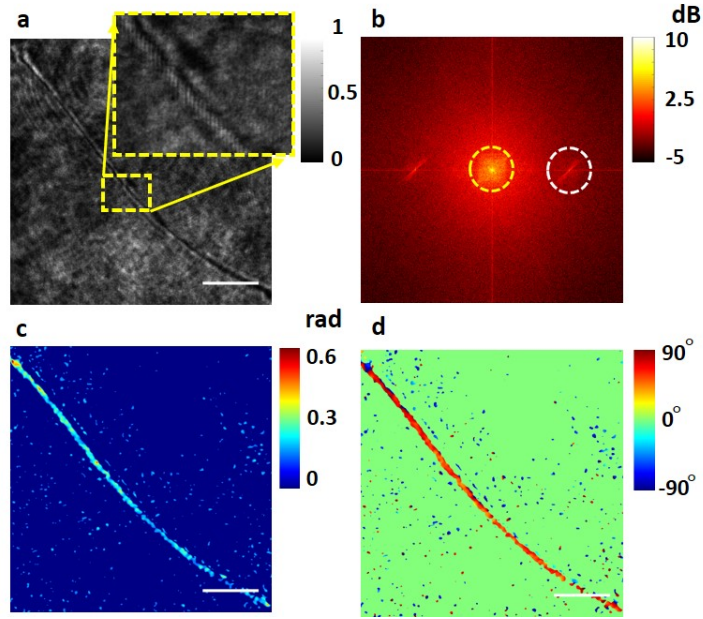

Figure 2. Demonstration of the polarization parameter retrieval algorithm. (a) Interferogram of a crystal fiber sample. The zoomed region denotes the fiber region with a high birefringence signal. (b) Logarithm map of the Fourier domain of (a), reported in decibel (dB), where the $0^{th}$ and $+1^{st}$ orders are labeled with yellow and white circles, respectively. (c) Quantitative map of the retardance distribution. (d) Quantitative map of the orientation angle distribution. The scale bar denotes 20 μm.

**RESULTS**

**Validation of PSIM's measurement accuracy using a rotating wave plate.** We measure the retardance and orientation angle of a zero-order quarter wave plate (QWP) at 532 nm to validate our methodology of PSIM. We rotate the fast axis of the QWP and take the interferograms at every 10°, as shown in Fig. 3(a). For each interferogram, we map the retardance and orientation angle distributions on a per pixel basis and average them over the entire field-of-view. The retardance and orientation angle mappings are calibrated with a zero-order QWP at a wavelength of 633 nm, as described in the Supporting Information. The average retardance and orientation angle, along with their standard deviations over the field-of-view, are shown in Figs. 3(c) and (e) as functions of the rotation angle. For the retardance, the value of a zero-order 532 nm QWP is 1.32 rad, and the mean value of the measured averaged retardance of all the rotation angles is 1.34 rad with standard deviation 0.01 rad. These two values represent the measurement accuracy for retardance. The averaged spatial standard deviation of the retardance value is 0.04 rad; it results from the speckle noise or nonuniformity of the background and is shown as error bars in Fig. 3(c). This value characterizes the spatial sensitivity of our PSIM system for mapping retardance.

The retrieved orientation angle tracks the physical rotation angle of the wave plate, as shown in Fig. 3(d). To avoid the influence of phase wrapping when processing the PSIM images, the Goldstein unwrapping algorithm is used to extend the period of the orientation angle from 90° to 180° [39]. The standard deviation of the differences between the measured and the actual values is 2.57°, the precision of the rotation stage

(Thorlabs, OCT-XYR1/M) is 1.00°. This value characterizes the measurement accuracy for the orientation angle. The spatial standard deviation of the retrieved orientation angle is 4.84°, characterizing the spatial sensitivity for the orientation angle and originating from the speckle-based spatial noise of the PSIM system.

**Validation of PSIM's measurement accuracy for complex birefringent structures using a rotating bovine tendon specimen.** The retardance and orientation angle of a bovine tendon specimen are quantitatively mapped and evaluated at different rotation angles, as shown in Figs. 3(b), (e) and (f). The bovine tendons are sliced into specimens of thickness 5 μm, sandwiched between a glass slide and a coverslip, and fixed on the rotation stage (Thorlabs, OCT-XYR1/M) with clippers. The bovine tendon contains abundant collagen fibers with large birefringence. The orientation of the fibers in the tendon are relatively uniform, except for small undulations. The tendon sample is an example of an anisotropic material with complex morphology, which makes it ideal for validating the measurement accuracy and imaging performance of PSIM. A bovine tendon specimen is imaged at rotation angles from 0° to 90° at increments of 10°. The retardance and orientation angle maps are retrieved at each rotation angle. Since the field-of-view at different rotation angles slightly shifts upon rotation, image registration method (reported in the Supporting Information) is used to find the most correlated regions in the retardance images at adjacent rotation angles. Maps of the polarization parameters of the most correlated regions at rotation angles of 0°, 10°, 20° and 30° are shown in Fig. 3(b), where each image combines the retardance images and the quiver plots of orientation angles. The direction of the rods in the quiver plots indicate that the retrieved orientation angle follows the undulation of the fibrous structures seen in the retardance images.

We further validate that the spatial average of the retrieved retardance maps remains constant, while the average values of the orientation angle maps match the rotation angles when we rotate the sample. The spatial averaged values and standard deviations of the retardance and orientation angle in the most correlated region found with our image registration algorithm are shown as functions of the rotation angle in Figs. 3(e) and (f). For the retardance, the mean of the spatially averaged values over all the rotation angles is 0.60 rad with standard deviation 0.01 rad, which indicates that the mean retardance changes negligibly as we rotate the sample stage. The spatial standard deviation of the retardance difference maps between adjacent rotation angles is 0.05 rad, indicating that the pixel mismatch caused by our image registration algorithm is negligible. For the orientation angle, the mean error between the averaged orientation angles and rotation angles is 2.39°, indicating that the orientation angle measurement with PSIM is accurate even for complex birefringent structures. The spatial standard deviation of the orientation angle difference between two adjacent rotation angles is 9.47°. This value is slightly larger than the spatial uncertainty of the retardance, partly due to the orientation angle being more sensitive to the mismatch of the tendon's fibrous structure. To avoid the inaccuracy introduced by the error of the mechanical rotation of the sample stage, the rotation angles are also determined by an image registration algorithm (details are reported in the Supporting Information).

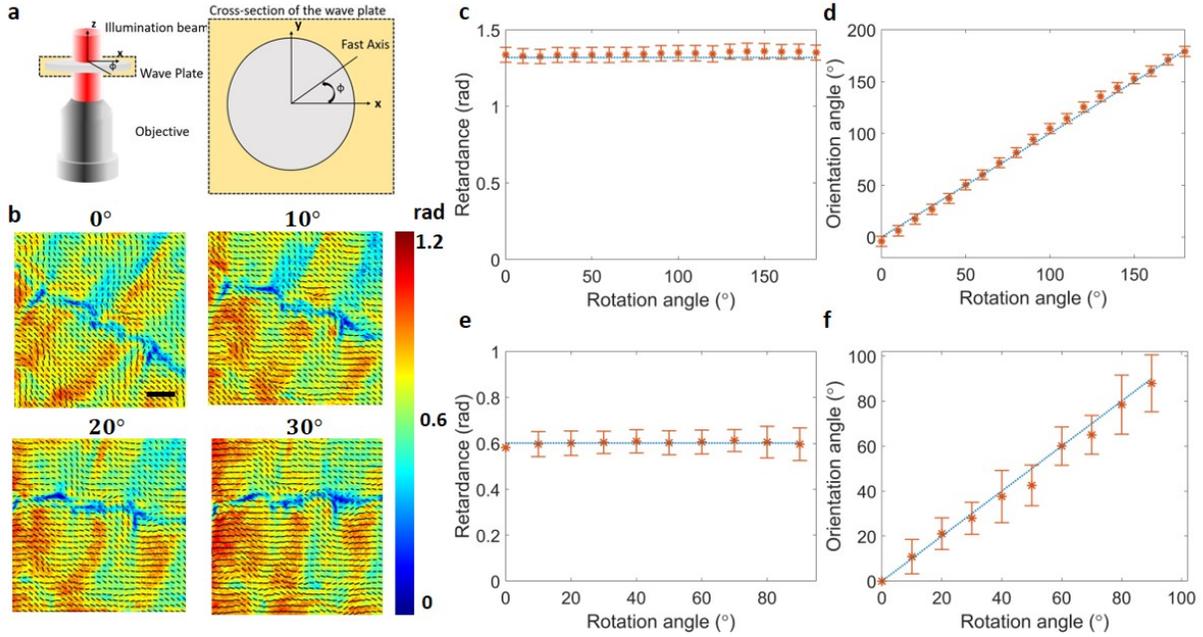

Figure 3. Validation of the imaging concept. (a) Scheme of the wave plate rotation. $\phi$ is the angle of the rotation, and we record an interferogram every 10°. (b) Combination of the retardance map and orientation angle's quiver plot of the bovine tendon specimen, for rotation angles of 0°, 10°, 20° and 30°. The scale bar denotes 25 µm. (c) and (d) Quantitative evaluation of the retardance and the orientation angle of a rotating QWP at 532 nm. The red asterisks denote the average values, the error bars denote the standard deviations over the field-of-view. The blue dashed lines show the nominal value of retardance (1.32 rad) and the values of rotation angles, respectively. (e) Averaged retardance and (f) orientation angle versus the rotation angle. The blue dashed lines denote the averaged value of retardance and the values of rotation angles, respectively.

**Quantitative polarization imaging of liquid crystal flows.** We apply PSIM to quantitatively image the flow of nematic liquid crystals in a microfluidic channel, as shown in Fig. 4(a). We control the initial conditions by injecting aqueous solutions of 13 wt% disodium cromoglycate (DSCG) into the microfluidic channel and allowing the DSCG solution to relax on rubbed surfaces where the directors are aligned along the flow direction (x-axis in Fig. 4(b)). We start the flow at a flow rate of 1 µl/min controlled by a syringe pump (Harvard PHD 2000). When the flow has reached a steady state, we image the sample in a 252×252 µm² region at an imaging speed of 506 fps.

The retardance and orientation angle maps depend on the flow behavior of the liquid crystal director field. The angle $\varphi$ denotes the in-plane orientation angle averaged along the z-direction, which describes the director orientation in the x-y plane, as shown in Fig. 4(a). The retardance $\Delta$ is the integrated phase difference between the projected extraordinary and ordinary axes of the liquid crystal.

We retrieve the retardance and orientation angle maps for each frame of the video (see Supporting Information for the video). The color map of the retardance and the quiver plot of the orientation angle are shown in Fig. 4(b), where the direction of flow is in the x-direction. The colors and the length of the rods represent the magnitude of retardance, the direction of the rods denotes the orientation angle. Note that the birefringence of a 13 wt% DSCG solution is $n_e - n_o \approx -0.015$ at a wavelength of 633 nm [40,41], where $n_e$ and $n_o$ are the extraordinary and ordinary refractive indices, respectively. The birefringence leads to a maximum retardance of ~ 0.98 ± 0.15 rad, well within the maximum measurable value π/2 of PSIM. Complex structures appear in the flow of DSCG: low retardance regions of well-defined size emerge in the steady-state flow.

**Tracking the temporal evolution of the retardance and orientation angle.** The spatial distribution of the retardance and orientation angle maps of DSCG solutions at shear rates below 10 s$^{-1}$ have recently been studied using an LC-Polscope [24]. Thanks to the large field-of-view and the high imaging speed of PSIM, we can quantify the spatiotemporal dynamics of the structures that emerge in the flow. To evaluate the dynamics, we focus on selected low-retardance regions of window size 100×100 pixels, as shown in Fig. 4(b), and track their evolution over time. We quantify the temporal evolution of the retardance and orientation angle by calculating the Pearson Correlation Coefficient (PCC) for time intervals of 2 ms, as shown in Fig. 4(c) (details of the analysis and the time-lapse videos of the polarization parameters of the selected regions are provided in the Supporting Information). The PCC values for both the retardance and the orientation angle monotonically decrease with time, indicating the motion of the structures. We report on the nature of the low retardance structures and their dynamics in a separate manuscript (in preparation).

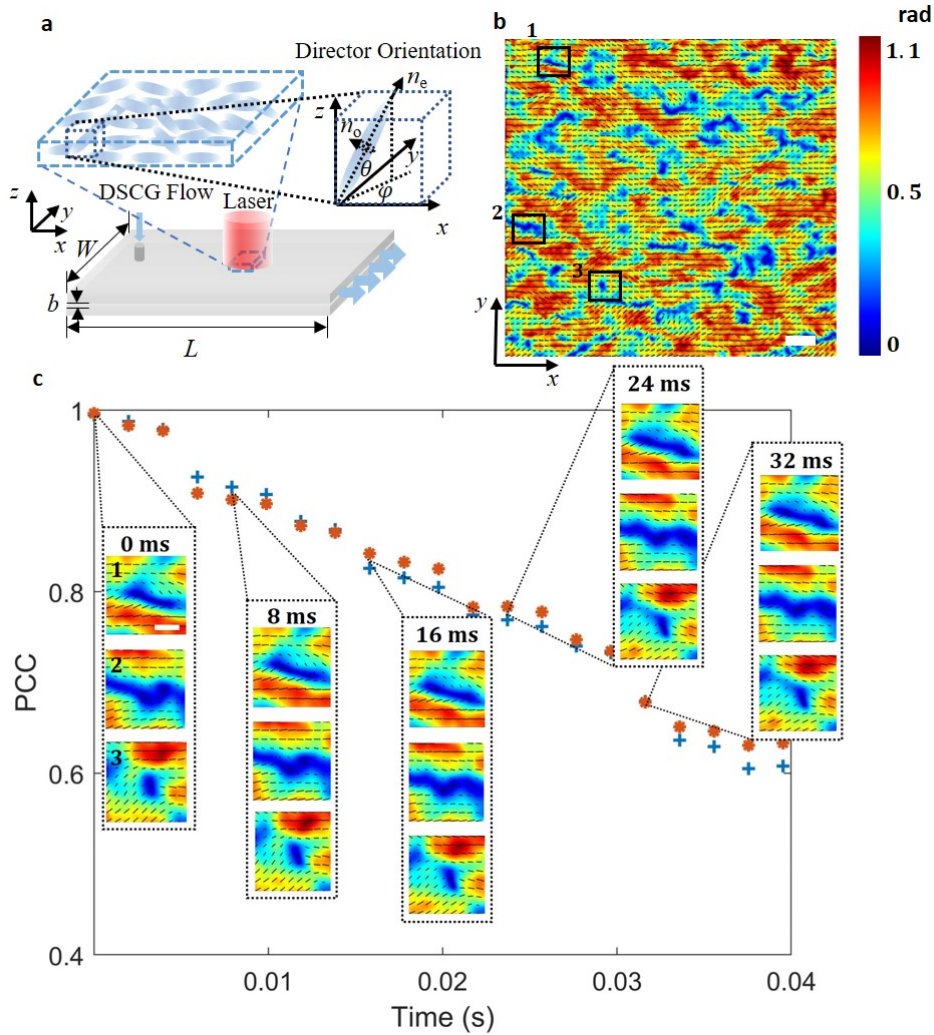

Figure 4. Temporal evolution of the retardance and orientation angle maps of a flowing liquid crystal. (a) Schematics of the microfluidic channel of length $L$ = 50 mm, $W$ = 15 mm, $b$ = 6.5 μm used to probe the flow of a nematic DSCG solution, and of the director orientation. (b) Retardance and orientation angle maps for a flow rate of 1 μl/min. The color and the length of the rods represent the magnitude of retardance, the direction of the rods denotes the orientation angle. The three regions of size 100×100 pixels discussed in (c) are indicated. The flow direction is from left to the right in the *x*-direction. (c) Temporal evolution of the Pearson Correlation Coefficient (PCC) of the retardance (blue plus signs) and orientation angle (red asterisks) maps, and snapshots of the three selected regions. The scale bars in (b) and (c) denote 20 μm and 10 μm , respectively.

**CONCLUSIONS**

We propose a novel imaging method termed polarized shearing interference microscopy (PSIM) that combines polarization microscopy and off-axis shearing interferometry to quantitatively image the structures and dynamics of birefringent materials. With the innovative PSIM system and a customized straightforward polarization parameter retrieval algorithm, we realize a simultaneous mapping of the polarization parameters of birefringent samples at imaging speeds only limited by the camera frame rate, while keeping a large field-of-view of up to 350×350 μm² and micrometer level spatial resolution of ~2 μm. We have validated that the measurement errors of retardance and orientation angle are 0.63% and 1.43% when imaging rotating wave plates, and we demonstrate that PSIM can also resolve the orientation change and the spatial undulations of bovine tendon fibers with high accuracy when imaging rotating specimens. The single-shot nature of PSIM allows us to quantify the fast dynamics of a nematic lyotropic chromonic liquid crystal flowing in a microfluidic channel and to track the emerging structures. Enabling high-speed quantitative polarization imaging, PSIM opens the path for quantifying and interpreting the dynamics of liquid crystals and other birefringent materials.

**METHODS**

**Derivation of the polarization retrieval algorithm.** We extract the retardance map from the amplitude of the retrieved complex field, and the orientation angle from the phase. The algorithm for retrieving the retardance and the orientation angle is derived with Jones calculus. The Jones matrix of the sample is:

$$J_{sample} = \begin{pmatrix} \cos\varphi & \sin\varphi \\ -\sin\varphi & \cos\varphi \end{pmatrix} \begin{pmatrix} e^{j\phi_e} & 0 \\ 0 & e^{j\phi_o} \end{pmatrix} \begin{pmatrix} \cos\varphi & -\sin\varphi \\ \sin\varphi & \cos\varphi \end{pmatrix}$$
$$= \begin{pmatrix} e^{i\phi_e}\cos^2\varphi + e^{i\phi_o}\sin^2\varphi & -\left(e^{i\phi_e} - e^{i\phi_o}\right)\sin\varphi\cos\varphi \\ -\left(e^{i\phi_e} - e^{i\phi_o}\right)\sin\varphi\cos\varphi & e^{i\phi_e}\sin^2\varphi + e^{i\phi_o}\cos^2\varphi \end{pmatrix}, \quad (3)$$

where $\phi_e$ is the phase delay in the extraordinary axis of the birefringent sample, $\phi_o$ is the phase delay in the ordinary axis, and $\varphi$ is the orientation angle. The retardance $\Delta$ is the difference between the phase delay in the extraordinary axis and that in the ordinary axis, $\Delta = \phi_e - \phi_o$. These parameters are mapped in real space, and the notations for Cartesian coordinates (*x*, *y*) are omitted for clarity. We utilize a linear polarizer (LP1) and a quarter wave plate (QWP1) to generate a left-handed circular polarization illumination:

$$E_{in} = \frac{1}{\sqrt{2}} \begin{pmatrix} 1 \\ i \end{pmatrix}. \quad (4)$$

After the beam is transmitted through the sample, the scattered light is collected by an objective and passes through another quarter wave plate (QWP2). The output field is:

$$E_{out} = e^{i\frac{\pi}{4}}\begin{pmatrix}1 & 0 \\ 0 & i\end{pmatrix}J_{sample}E_{in}$$
$$= \frac{1}{\sqrt{2}}e^{i\frac{\pi}{4}}\begin{pmatrix}e^{i\phi_e}\left(\cos^2\varphi - i\sin\varphi\cos\varphi\right) + e^{i\phi_o}\left(\sin^2\varphi + i\sin\varphi\cos\varphi\right) \\ e^{i\phi_e}\left(-\sin^2\varphi - i\sin\varphi\cos\varphi\right) + e^{i\phi_o}\left(-\cos^2\varphi + i\sin\varphi\cos\varphi\right)\end{pmatrix}.$$

(5)

Subsequently, the light is separated by a diffraction grating. Only the +1 and –1 orders of the light pass through the Fourier plane. After a polarizer with orientation direction of 45° to the slow axis of QWP2, an output field is produced:

$$E_{out,45°} = \frac{1}{2}e^{i\frac{\pi}{4}}\left(e^{i\phi_e} - e^{i\phi_o}\right)\exp(-i2\varphi) = \sin\frac{\Delta}{2}\exp\left[i\left(\frac{\phi_e + \phi_o}{2} - 2\varphi + \frac{3\pi}{4}\right)\right].$$

(6)

Note that the retardance is only contained in the amplitude part and the orientation angle is only contained in the phase part. However, the presence of an average phase delay ($\frac{\phi_e + \phi_o}{2}$) prevents the retrieval of the orientation angle. The output electric field for a linear polarizer set to –45° to the slow axis of QWP2 is expressed as:

$$E_{out,-45°} = \frac{1}{2}e^{i\frac{\pi}{4}}\left(e^{i\phi_e} + e^{i\phi_o}\right) = \cos\frac{\Delta}{2}\exp\left[i\left(\frac{\phi_e + \phi_o}{2} + \frac{\pi}{4}\right)\right].$$

(7)

The orientation angle contained in the phase part disappears and only the average phase delay remains. This offers a strategy to cancel out the average phase delay. Two perpendicularly oriented polarizers are placed on the Fourier plane to generate these two output fields simultaneously: one is in the +1st order, the other is in the –1st order:

$$E_{+1} = \sin\frac{\Delta}{2}\exp\left[i\left(\frac{\phi_e + \phi_o}{2} - 2\varphi + \frac{3\pi}{4} + \frac{kx}{2}\right)\right],$$

(8)

and

$$E_{-1} = \cos\frac{\Delta}{2}\exp\left[i\left(\frac{\phi_e + \phi_o}{2} + \frac{\pi}{4} - \frac{kx}{2}\right)\right].$$

(9)

$kx$ denotes the spatial modulation caused by the separation of the diffraction grating. A second polarizer (LP2) with orientation 45° to both polarizers on the Fourier plane is used to produce interference, and an interferogram is recorded by a CMOS camera:

$$I = \langle(E_{+1} + E_{-1})(E_{+1} + E_{-1})*\rangle$$
$$= |E_{+1}|^2 + |E_{-1}|^2 + \langle E_{+1}E_{-1}^*\rangle + \langle E_{-1}E_{+1}^*\rangle$$
$$= \sin^2\frac{\Delta}{2} + \cos^2\frac{\Delta}{2} + 2\sin\frac{\Delta}{2}\cos\frac{\Delta}{2}\cos\left[\left(\frac{\phi_e + \phi_o}{2} - 2\varphi + \frac{3\pi}{4}\right) - \left(\frac{\phi_e + \phi_o}{2} + \frac{\pi}{4}\right) + kx\right]$$
$$= 1 + \sin\Delta\sin(2\varphi - kx).$$

(10)

By retrieving the complex field in the alternating current (AC) term, we can calculate the retardance and orientation angle distribution, Eqns. (1) and (2).

**Sample Preparation.** Orange II is purchased from Sigma-Aldrich and aqueous solutions with weight concentration $c$ = 35.0 wt% are prepared in the nematic phase.

Bovine tendon is purchased at a local market. After cutting the tendon samples with a scalpel they are stored in 10% formalin (Sigma-Aldrich) for at least 72 hours. Subsequently, the samples are dehydrated in ethanol and embedded in paraffin to fix the thin sections to be imaged.

Disodium cromoglycate (DSCG) is purchased from TCI America (purity > 98.0 %). The DSCG is dissolved in deionized water at a concentration of $c$ = 13.0 wt% [42]. The linear microfluidic channel consists of two rubbed rectangular glass plates separated by 6.5 ± 1 μm spacers (Specac, MY SPR RECT 0.006 mm OMNI). We use diamond particles of diameter ~ 50 nm to rub both glass plates along the channel length direction, which induces uniform planar alignment.

## ACKNOWLEDGMENTS

B.G., Z.Y. and P.T.C.S. acknowledge support from the National Institutes of Health (NIH) 5-P41-EB015871-27, 5R21NS091982-02, and the Hamamatsu Corporation. B.G. and P.T.C.S. acknowledge support from the Singapore–MIT Alliance for Research and Technology (SMART) Center, Critical Analytics for Manufacturing Personalized-Medicine IRG. B.G. acknowledges support from MathWorks Fellowship. Q.Z. and I.B. acknowledge support from the MIT Research Support Committee. Z.Y. and P.T.C.S. acknowledge support from NIH R01DA045549 and R21GM140613-02. Rui Z. acknowledges support from the Hong Kong RGC grant no. 26302320. R. Z. acknowledges the Croucher Foundation (Award Number: CM/CT/CF/CIA/0688/19ay). C.Y.D., J.T.L., P.H.T. and C.W.C. acknowledge 107-2112-M-002-023-MY3, Ministry of Science and Technology, Taiwan, R.O.C.